\documentclass{ifacconf}

\usepackage{amsmath}
\allowdisplaybreaks[4]

\usepackage{epsfig,tabularx,amsfonts,amssymb,amsmath}
\usepackage{epstopdf}
\usepackage{graphics}
\usepackage{epic,eepic,epsfig}
\usepackage{mathrsfs}
\usepackage{calc,ifthen}
\usepackage{multirow}
\usepackage{bm}
\usepackage{float}
\usepackage{pstricks}
\usepackage{url}
\usepackage{setspace}
\usepackage{psfrag}
\usepackage{pstool}
\usepackage{changes}
\usepackage{commath}
\usepackage[normalem]{ulem}

\graphicspath{{figures/}}  

\usepackage{color}

\newtheorem{assumption}{Assumption}

\newcommand{\mynote}[1]{}
\newcommand{\newnote}[1]{}
\newcommand{\cancelled}[1]

\usepackage{natbib}      
\begin{document}
\begin{frontmatter}

\title{Economic Potential for Hybrid Electric Vehicles in Urban Signal-free Intersections with Decentralized MPC}


\author[First]{Kai Tang}
\author[Second]{Weijie Wang}
\author[Second]{Xiao Pan}
\author[Third]{Boli Chen}
\author[Second]{Simos A. Evangelou}

\address[First]{Department of Electrical and Electronic Engineering, University of Hong Kong, Hong Kong, (e-mail: tangkai@eee.hku.hk).}
\address[Second]{Department of Electrical and Electronic Engineering, Imperial College London, UK, (e-mail: weijie.wang20@imperial.ac.uk, xiao.pan17@imperial.ac.uk, s.evangelou@imperial.ac.uk).}
\address[Third]{Department of Electronic and Electrical, University College London, UK, (e-mail: boli.chen@ucl.ac.uk)}

\begin{abstract}                
The development of electric and connected vehicles as well as automated driving technologies are key towards the smart city, with convenient urban mobility and high energy economy performance. 
However, the global rise in electricity price provokes renewed interest on CAVs with hybrid electric powertrains rather than considering battery electric powertrains. 
This paper provides a decentralized coordination strategy for a group of connected and autonomous vehicles (CAVs) with a series hybrid electric (sHEV) powertrain at signal-free intersections. 
The problem is formulated as a convex form with suitable relaxation and approximation of the powertrain model and solved by decentralized model predictive control, which is able to ensure a rapid search and unique solution in real time.
Numerical examples validate the effectiveness of the proposed methods concerning physical and safety constraints.
By utilizing the petrol fuel and battery charging prices over the last year, the performance of the proposed approach is evaluated against the optimal results produced by two benchmark solutions, conventional vehicles (CVs) and battery electric vehicles (BEVs). The comparison results show that the traveling cost of sHEVs approaches and even under some circumstances reaches the same level as for BEVs, which indicates the importance of hybridization, particularly under the current rising electricity price situation.
\end{abstract}


\end{frontmatter}

\section{Introduction}
Rapid urbanization causes an increasing number of vehicles and intensifies traffic congestion, especially in urban districts. This phenomenon can lead to many problems, such as potential safety hazards and increased energy and time consumption, particularly in urban intersections, where the interaction between different traffic flows is extremely complex. Although some research focuses on designing novel traffic light systems to alleviate these problems \citep{yang2017cooperative, Xiang2018Signal}, the efficiency of the traffic flow at an intersection and the energy consumption is still not fully optimized~\citep{namazi2019intelligent}. To maximize urban transportation efficiency, a smoother and environmentally friendly scheme needs to be proposed. Nowadays, the development of information technology enables vehicle-to-everything (V2X) communication for connected and autonomous vehicles (CAVs) to be deployed on real-world road networks~\citep{zhao2022enhanced}. Therefore, it is an attainable goal to realize a signal-free intersection control to optimally coordinate a group of CAVs crossing the intersection for safety, mobility, and energy efficiency objectives~\citep{zhong2020autonomous}.

There have been numerous efforts of autonomous intersection control algorithms designed based on CAVs reported in the literature, which can be divided into two categories, centralized and decentralized, respectively~\citep{vcakija2019autonomous, Gholamhosseinian2022}. 
The centralized framework coordinates the vehicles' movement using one central intersection controller (IC)~\citep{add_centralized, mihaly2020model, liu2020high}, while the decentralized framework relies on local controllers on each vehicle~\citep{chalaki2021priority,chalaki2022optimal,hadjigeorgiou2022real}. Some existing literature focuses on battery electrical vehicles (BEVs) as the target CAVs for urban intersection crossing problems, instead of conventional vehicles (CVs), as BEVs are the current trend of vehicle development offering the advantage of reduced operating cost and waste gas emissions, although they also suffer from low power and long charging time~\citep{Hult:ecc19, pan2022TCST, pan2022optimal}. In \citep{Hult:ecc19}, the energy consumption is modeled by electric motors efficiency maps and optimized by an economic model predictive control (MPC) method.
Recent research efforts in \citep{pan2022TCST,pan2022optimal} develop hierarchical and convex optimization approaches to find the optimal trade-offs of battery energy consumption and travel time.

However, the global electricity price has increased sharply recently. For instance, the UK electricity price has risen by 64\% from about $226.59$ (GBP/MWh) to $372.27$ (GBP/MWh) during the last 12 months, compared with the lower rate increase of fuel price by 18\% from 138.64 (pence/liter) to 163.62 (pence/liter)~\citep{ukelectricity, ukgasoline}. This phenomenon significantly influences the operational cost of the BEV, and it may promote hybrid electric vehicles (HEVs), which are powered by hybrid energy sources. This paper aims to investigate the energy-saving potential of HEVs in comparison with BEVs in the context of a road intersection, which is the main bottleneck of urban traffic. The work focuses on the series HEV (sHEV) architecture, which is a common arrangement for modern HEVs and involves a number of products in the market, such as the Nissan Note e-Power, and numerous other extended-range electric vehicles \cite{chen2019series}.
Some recent literature that considers sHEVs as a CAV option for urban signalized intersection crossing problems is presented in~\cite{shevTANG2021, shevJian2021}. In \cite{shevTANG2021}, a multi-objective hierarchical optimal strategy is proposed to optimize fuel consumption and riding comfort in signalized intersections. A two-level cooperative control method is designed in \cite{shevJian2021} to improve the travel time and energy consumption of sHEV in a signalized intersection.
However, to the best knowledge of the authors, existing researches lack investigating convex optimization-based autonomous intersection management approaches, particularly for sHEVs. 
This paper works on bridging the gap by proposing a convex optimization framework of sHEVs for the signal-free intersection crossing problem. The resulting problem is then solved by a decentralized model predictive control (DMPC) approach, rather than centralized MPC, that strikes the trade-off between optimality and computational complexity~\citep{pan2022TCST}. The contributions of this paper are therefore: 1) development of a new convex modeling framework for eco-driving control of an sHEV, 
2) utilization of the contributed convex modeling framework of a single sHEV to develop a decentralized convex model predictive control for autonomous intersection coordination of a fleet of sHEVs, and
3) numerical investigation of the energy-saving potential for sHEVs in urban signal-free intersections; in the context of the UK's electricity price, the running cost of the sHEV in the urban intersection is compared with the CV and BEV scenarios.

The rest of this paper is organized as follows: Section~\ref{sec:problemstatement} provides the signal-free intersection and sHEVs powertrain models along with the corresponding approximation and convexification. The decentralized model predictive control problem is formulated in Section~\ref{sec:controlframework}. Numerical results and analyses are provided in Section~\ref{sec:simulation}. Finally, concluding remarks are given in Section~\ref{sec:conclusions}.

\section{Problem Statement}
\label{sec:problemstatement}

\subsection{Convex Modeling of Intersection Crossing Problem}
\begin{figure}[htp!]
    \centering
    \includegraphics[width=.60\columnwidth]{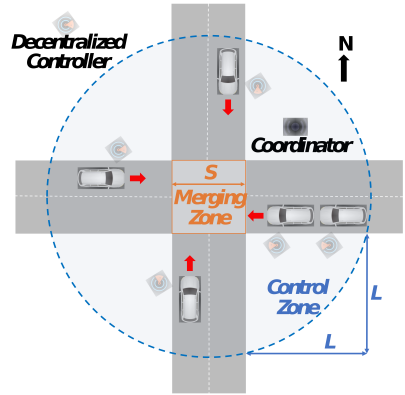}\\ [-2ex]
    \caption{The schematic of a decentralized signal-free intersection model with CAVs and a coordinator.}
    \label{fig:intersection}
\end{figure}

This work focuses on the decentralized intersection problem where a group of CAVs approaches a signal-free intersection with two perpendicular roads, north-south, and east-west, respectively, as shown in Fig.~\ref{fig:intersection}.
Each road has two lanes serving vehicles traveling in opposite directions. It is assumed that there are no other non-autonomous road users (e.g., human-driven vehicles, cyclists, and pedestrians). As shown in Fig.~\ref{fig:intersection}, the squared area in the central part is defined as \textit{Merging Zone} (MZ) of length $S$, where vehicles merge from four directions, and the potential of lateral collisions exists. 
The circle area illustrated by the blue line outside the MZ is defined as \textit{Control Zone} (CZ) with distance $L$ from the entry of the CZ to the entry of the MZ. Note that the physical length of the MZ is much smaller than the sensing range of the wireless communication devices, i.e., $S\!<\!L$. The communication among CAVs is through an intersection coordinator, which acts as a relay to streamline the communication network without making any control decisions. The control actions are individually determined by the local onboard controller of each vehicle. For simplicity, it is assumed that the roads are flat, and all CAVs maintain their initial directions after exiting the MZ. 

Next, let us define $N(t)\! \in\! \mathbb{N}_{>0}$ as the whole number of CAVs in the CZ at a given time $t\!\in\! \mathbb{R}_{>0}$. 
The crossing order of CAVs is designated by a set  $\mathcal{N}(t)\!=\!\{1,2,\ldots,N(t)\}$. 
Note that the crossing order sequence in this work is considered to be predefined for a simplicity purpose of illustration of the developed convex modeling framework, and it can be formulated as an online optimization problem to further optimize the traffic performance~\citep{Meng:18,hadjigeorgiou2022real,pan2022optimal}.
The following definitions are presented for further discussion of collision avoidance constraints.
Given an arbitrary $i$th CAV, $i\!\in\!\mathcal{N}$, the CAVs entering the MZ before the $i$th CAV can be divided into three categories: 1) $\mathcal{C}_{i}$ includes the CAVs traveling towards the same direction as the $i$th CAV; 2) $\mathcal{L}_i$ includes the CAVs traveling towards the perpendicular direction of the $i$th CAV; 3) $\mathcal{O}_i$ includes the CAVs traveling towards the opposite direction of the $i$th CAV.

This research focuses on developing a coordination scheme in the space domain, which has been shown to have advantages of (a) avoiding free end-time optimization problems in the time domain, and (b) achieving convex optimization problem formulation~\citep{pan2022optimal}. 
The travel time of each CAV $i$ in space domain, in which the variable of traveled distance is $s$, can be easily formulated as a state variable with a linear dynamics equation and a convex path constraint: 
\begin{subequations}\label{eq:dt}
\begin{align}
&  \frac{d}{ds}t_i(s)  = \zeta_i(s), \label{eq:dt_equality}\\
& \zeta_i(s)\geq \frac{1}{\sqrt{2E_i(s)/m_i}}, \label{eq:dt_path}
\end{align}
\end{subequations}
where $E_i(s)\!=\!\frac{1}{2}m_i v_i^2(s)$ is the kinetic energy, $m_i$ is the vehicle mass, and $\zeta_i$ is an optimization auxiliary variable. 
Note that the validity
of the final solution relies on the tightness of \eqref{eq:dt_path}, and an analogical proof of the tightness can be found in~\cite{pan2022TCST}.
As such, the required travel time of each CAV $i$ to cross the intersection is:
\begin{equation}\label{eq:J_ti}
J_{t,i}= t_i(L+S)-t_i(0),
\end{equation}
where $t_i(0)$ is the arrival time of CAV $i$ at the CZ, and $L\!+\!S$ is the total traveled distance, which represents the complete task of each vehicle crossing the intersection. Herein, $E_i(s)$ is introduced intentionally for modeling the motion dynamics instead of $v_i$ in order to cancel the nonlinearity due to the air drag, as shown in the following equation \eqref{eq:space_dv}. Considering $E_i(s)$, the longitudinal dynamics of vehicle $i$ can be described by:
\begin{equation}\label{eq:space_dv}  
\frac{d}{ds} E_i(s) = F_{w,i}(s) -F_{r,i}-\frac{2f_{d}}{m_i} E_{i}(s),
\end{equation}
where $F_{r,i}\!=\!f_{r} {m_i} g$, with coefficient $f_{r}\!=\!0.01$, is the rolling resistance force and $f_{d}\!=\!0.47$ is the coefficient of air drag resistance. $F_{w,i}(s)=F_{t,i}(s)+F_{b,i}(s)$ is the total force acting on the wheels, with $F_{t,i}(s)$ the powertrain driving force and $F_{b,i}(s)$ the mechanical braking force, respectively. Note that $F_{b,i}(s)$ is constrained by:
\begin{equation}\label{eq:Fb_limits}
 m_ia_{\min}-F_{{t,i}_{\min}}\leq F_{b,i}(s) \leq0,
\end{equation}
where $a_{\min}\!=\!-6.5~m/s^{2}$ is the minimum allowed acceleration due to the tire friction limits, and $F_{{t,i}_{\min}}$ is the minimum traction (maximum regenerative) force provided by the powertrain electric sources.

To avoid vehicle collisions inside the CZ and to maintain a smooth traffic flow, lateral and rear-end collision  avoidance constraints and speed limits are introduced as:
\begin{align}
& t_i(s)\!-\!t_h(s) \geq \beta(s),\,\,\forall h\in\mathcal{C}_i,\,\,h<i,\label{eq:TTCapprox}\\
 & t_i(L)\geq t_q(L+S)\,,\,\,\, \forall q\in\mathcal{L}_i\,,\,\,q<i,\label{eq:lateral}\\
&\frac{1}{2}m_iv_{\min}^2 \leq \,E_i(s)\, \leq \frac{1}{2}m_iv_{\max}^2, \label{eq:Bound_E_i}
\end{align}
where
\begin{equation}
    \beta(s) =\max\left(\frac{\sqrt{2E_i(s)/m_i}\!-\!\sqrt{2E_h(s)/m_h}}{|a_{\min}|},\, t_{\delta}\right)\,,
\end{equation}
$t_{\delta}$ is a short time constant to enforce a safety margin invariably~\citep{pan2022optimal}, $v_{\min}\!=\!0.1\,$m/s is set to a sufficiently small positive constant to avoid singularity issues that would appear in \eqref{eq:dt} when $v_i{=}0$, and $v_{\max}\!=\!15\,$m/s is determined as the maximum velocity based on the infrastructure constraints and traffic regulations~\citep{hadjigeorgiou2022real}. 
The rear-end collision avoidance constraint \eqref{eq:TTCapprox} enforces a minimum time gap between CAV $i$ and CAV $h$, which is the vehicle immediately ahead of CAV $i$. Note that the convexification of \eqref{eq:TTCapprox} is achieved by an approximation of the nonlinear term $\sqrt{2E_i(s)/m_i}$ to $f^*(E_i(s))\!=\!a_{0,i}^*\!+\!a_{1,i}^*E_i(s)$, where $a_0^*$ and $a_1^*$  are obtained by a constrained least-squares optimization to maximize the feasibility while preserving the convexity~\citep{pan2022optimal}.
Lateral collisions constraint \eqref{eq:lateral} guarantees that CAV $i$ enters the MZ only after CAV $q$ has left the MZ, with CAV $q$ being the last vehicle entering the MZ in set $\mathcal{L}_i$.

For any CAV $p \in \mathcal{O}_i$, there is no interference between CAVs $p$ and $i$ inside the CZ. Hence, for CAV $p$ which is the last CAV entering the MZ  before CAV $i$ in the opposite direction, only the following constraint \eqref{eq:opposite} is required to fulfill the crossing order:
\begin{equation}\label{eq:opposite}
t_i(L)\!>\!t_p(L),\,\,\,t_i(L\!+\!S)\!>\!t_p(L\!+\!S),\, \,\,\forall p\in\mathcal{O}_i,\,\,p<i.
\end{equation}

Finally, each CAVs is requested to leave the MZ at a desired terminal speed $\bar{v}_i \in [v_{\min},v_{\max}]$,
\begin{equation}\label{eq:terminalv} 
E_i(L+S) = \frac{1}{2}m_i\bar{v}_i^2.
\end{equation}
Note the terminal speed $\bar{v}_i$ can be chosen arbitrarily within $ [v_{\min},v_{\max}]$, which will be specified later in Section~.\ref{sec:simulation}.

The following assumptions are also needed to fulfill the modeling of the autonomous intersection problem.
\begin{assumption}\label{ass:delay}
	The information of each CAV, such as position and velocity, can be measured and transferred between the coordinator without errors and delays.
\end{assumption} 
\begin{assumption}\label{ass:entrytime}
Each CAV enters the CZ at a different time, i.e., $t_i(0)\!\ne\!t_z(0), \,i\!\ne\!z$, $i,z\in\{1,2,\cdots,N(t)\}$. 
\end{assumption}
\begin{assumption}\label{ass:initial}
For each CAV $i$, constraints \eqref{eq:TTCapprox}, and \eqref{eq:Bound_E_i} are inactive at $t_i(0)$.
\end{assumption}
Assumption \ref{ass:delay} may not be valid for practical situations. In that case, it can be relaxed by using a worst-case analysis as long as the uncertainties in measurement and communication are bounded.
Assumption~\ref{ass:entrytime}-\ref{ass:initial} are intended to ensure the feasibility of the initial states and the constraints \eqref{eq:TTCapprox}, \eqref{eq:lateral}, \eqref{eq:opposite} with predefined given order $\mathcal{N}(t)$.

\subsection{Convex Modeling of Series Hybrid Electric Vehicle}\label{subsec:convexmodel}
This section provides a convex dynamic model of an sHEV whose powertrain and dynamics are relaxed and conservatively convexified, which is one of the main contributions of this work. 
The main parameters of the sHEV are listed in Table.~\ref{tab:HEVdata}~\citep{chen2019series}, which emulate a non-plug-in sHEV. 
\begin{table}[!ht]
\centering 
\caption{Main Parameters of Series Hybrid Electric Vehicle Model} \vspace{-1ex}
\label{tab:HEVdata} 
\begin{tabular*}{1\columnwidth}{l @{\extracolsep{\fill}} c@{\extracolsep{\fill}}l}
\hline
\hline
 symbol & value & description \\
 \hline
 $m_i$ & 1200kg\ & vehicle mass  \\
 $m_{f0,i}$ & 0.061\,g/s & idling fuel consumption rate \\
 $\alpha_{PS}$ & 0.059 & scale factor  \\
 $a_{\text{SOC}}^*$ & -1.944$\times 10^{-7}$ & SOC fitting parameter \\
 $a_{\text{0},i}^*$ & -8.5034$\times 10^{-5}$ & velocity fitting parameter \\
 $a_{\text{1},i}^*$ & 4.9 & velocity fitting parameter \\
 $P_{PS_{\max}}$ & 75kW & maximum PS output power \\
 $P_{SS_{\min}}$ & -15kW & minimum SS output power \\
 $P_{SS_{\max}}$ & 30kW & maximum SS output power \\
\hline
\hline
\end{tabular*}
\end{table}

\begin{figure}[htp!]
    \centering
    \includegraphics[width=.95\columnwidth]{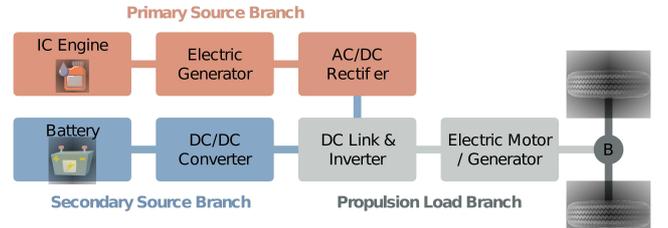}\\ [-2ex]
    \caption{Powertrain configuration for sHEVs.}
    \label{fig:SHEV}
\end{figure}

The powertrain model of the sHEV utilized in this work is sketched in Fig.~\ref{fig:SHEV}. As it can be seen, the power outputs from the \textit{Primary Source} (PS) and the \textit{Secondary Source} (SS) are jointly connected to the DC-link, which delivers driving power at the wheels included in the \textit{Propulsion Load} (PL) branch. Mechanical brakes can be actuated to decelerate the vehicle. Energy regeneration is also possible during deceleration conveying braking power through the transmission up to the battery. Moreover, the SS can be recharged by using a fraction of the PS power.

The PS branch comprises an internal combustion engine, a permanent magnet
synchronous generator, and an AC-DC rectifier connected in series. The mechanical separation of the PS from the driving wheels allows the PS to be operated along its optimal efficiency trajectory of operating points. In this context, the fuel mass rate of the PS can be approximated as a linear function of PS output power, $P_{PS}$~\citep{chen2019series}. Thus, the dynamic of the fuel mass consumption of an sHEV in the space domain is given by:
\begin{equation}\label{eq:HEV_m_f}
   \begin{aligned}
     \frac{d}{ds} m_{f,i} \!=\! \frac{m_{f0,i} \!+\! \alpha_{PS} P_{PS,i}(s)}{v_i(s)}
     \!=\!m_{f0,i}\zeta_i(s)\! +\! \alpha_{PS} F_{PS,i}(s),
   \end{aligned}
\end{equation}
where $m_{f0,i}$ is the idling fuel consumption rate,  $\alpha_{PS}$ is the scale factor between the fuel consumption rate and the PS output power, and $F_{PS,i}\!=\!{P_{PS,i}}/{v_i}$, proposed as a virtual force of PS, is constrained by the powertrain physical capability, namely $P_{PS_{\max}}\!=\!75$kW, such that:
\begin{equation}\label{eq:HEV_FPS_bound}
    0 \leq F_{PS,i}(s) \leq P_{PS_{\max}}\zeta_i(s).
\end{equation}

The SS branch consists of a battery and a DC/DC converter. The battery is modeled as a series connection of an ideal voltage source (of open-circuit voltage $V_{oc}$) and an ohmic resistance (internal resistance $R_b$)~\citep{zhou2016pseudospectral}, in which the battery state-of-charge (SOC) represents the only state variable, governed by:
\begin{equation}\label{eq:SOC}
    \frac{d}{d t} \mathrm{SOC}_i(t)=\frac{-V_{o c}+\sqrt{V_{oc}^{2}-4 P_{SS,i}(t) R_{b} }}{2 R_{b} Q_{\max }},
\end{equation}
where the operation of the battery SOC is constrained to avoid deep charging/depleting:
\begin{equation}\label{eq:SOC_limits}
    \mathrm{SOC}_{\min}\leq\mathrm{SOC}_i(t)\leq\mathrm{SOC}_{\max}.
\end{equation}

\begin{figure}[htp!]
    \centering
    \includegraphics[width=.95\columnwidth]{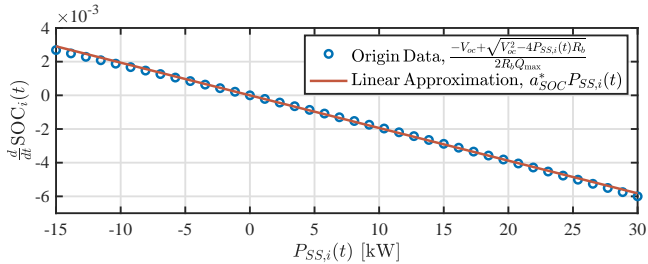}\\ [-2ex]
    \caption{Linear regression function \eqref{eq:SOC_approximate_timeDomain} of the derivative of the battery SOC, $\frac{d}{dt}\text{SOC}_i$, with respect to the SS output power, $P_{SS,i}$, with R-square of 99.89\%.}
    \label{fig:SOC_fitting}
\end{figure}

The nonlinear terms existing in the right-hand side of \eqref{eq:SOC} can be approximated as a linear proportional function (see in Fig.~\ref{fig:SOC_fitting}) 
for convex formulation.
As such, this function should pass through the origin to preserve the property of charging/depleting when $P_{SS}$ is negative/positive. The proposed formulation is given by:
\begin{equation}\label{eq:SOC_approximate_timeDomain}
    \frac{d}{d t} \mathrm{SOC}_i(t)=a_{\text{SOC}}^* P_{SS,i}(t), 
\end{equation}
where $a_{\text{SOC}}^*\!=\!-1.940\!\times\! 10^{-7}$ is the fitting parameter obtained by the constrained least-squares method subject to the physical limits of the SS branch:
$$
\begin{aligned} 
& \min \limits_{a_{\text{SOC}}}\left[a_{\text{SOC}} P_{SS,i}(t)-\frac{-V_{o c}+\sqrt{V_{oc}^{2}-4 P_{SS,i}(t) R_{b} }}{2 R_{b} Q_{\max }}\right]^{2} \\
& \text { s.t.: }   P_{{SS}_{\max}} \leq P_{SS,i}(t)\leq P_{{SS}_{\min}} \\
\end{aligned}
$$
Therefore, the dynamics of SOC \eqref{eq:SOC} can be reformulated as a linear function in the space domain:
\begin{equation}\label{eq:SOC_approximate}
    \frac{d}{d s} \mathrm{SOC}_i(s)=a_{\text{SOC}}^* \frac{P_{SS,i}(s)}{v_i(s)}=a_{\text{SOC}}^* F_{SS,i}(s), 
\end{equation}
where $F_{SS,i}\!=\!{P_{SS,i}}/{v_i}$ proposed as a virtual force of the SS branch is constrained by the following equation:
\begin{equation}\label{eq:HEV_FSS_bound}
 P_{SS_{\min}}\,\zeta_i(s) \leq F_{SS,i}(s) \leq P_{SS_{\max}}\,\zeta_i(s),
\end{equation}
where $P_{SS_{\min}}$ and $P_{SS_{\max}}$ are the maximum SS branch power during depleting/charging.

In the sHEV model architecture, the total powertrain force defined in the longitudinal dynamics \eqref{eq:space_dv} can be specified as below:
\begin{equation}\label{eq:HEV_Ft_Fps_Fss}
F_{t,i}(s)=F_{PS,i}(s)+F_{SS,i}(s).
\end{equation}

Thus, by collecting \eqref{eq:dt_equality}, \eqref{eq:space_dv}, \eqref{eq:HEV_m_f}, and \eqref{eq:SOC_approximate}, the overall dynamics of each sHEV $\dot{{x}_i}\!=\!f_i({x}_i,{u}_i)$ in the autonomous intersection problem can be rearranged as a state-space form equation with states ${x}_i\!=\! [t_i\,\,E_i\,\,\mathrm{SOC_i}\,\,m_{f,i}]^{\top}$ and control inputs ${u}_i\!=\! [F_{PS,i}\,\,F_{SS,i}\,\,F_{b,i}\,\,\zeta_{i}]^{\top}$:
\begin{equation}
\begin{aligned}
    &\frac{d}{ds}{x}_i(s)=A_i{x}_i(s)+B_i{u}_i(s)+B_{c,i}\,,\label{eq:HEV_state}\\
    &A_i\!=\!\begin{bmatrix}
    0&0&0&0\\
    0&-\frac{2f_{d}}{m_i}&0&0\\
    0&0&0&0\\
    0&0&0&0
    \end{bmatrix}\!,\, B_i\!=\!\begin{bmatrix}
    0&0&0&1\\
    1&1&1&0\\
    0&a_{\text{SOC}}^*&0&0\\
    \alpha_{PS}&0&0&m_{f0,{i}}
    \end{bmatrix}\!,\\ 
    & B_{c,i}\!=\!\begin{bmatrix}
    0&-F_{r,i} & 0&0
    \end{bmatrix}^{\top}\!. 
\end{aligned}
\end{equation}

\section{Decentralized Autonomous Intersection Control Framework}
\label{sec:controlframework}

This section provides an introduction to the autonomous intersection crossing control approach via the decentralized model predictive control (DMPC) technique with a prediction horizon $N_p\!\in\!\mathbb{N}_{>0}$. Hence, it is necessary to formulate the intersection control problem in a discretized form with a sampling distance interval $\Delta s\!\in\! \mathbb{R}_{>0}$.
Meanwhile, let us define that $L\!+\!S\!=\!\alpha\Delta s$, $L\!=\!\alpha_1\Delta s,\,S\!=\!\alpha_2\Delta s$. 
The discretized dynamics \eqref{eq:HEV_state} of CAV $i\!\in\! \mathcal{N}$ in the space domain is given by:
\begin{equation}\label{eq:dynamics_discretize}
    {x}_i(k+1)=A_i{x}_i(k)+B_i{u}_i(k)+B_{c,i},
\end{equation}

The decentralized system relies on the information exchange between the intersection coordinator and the CAVs. For an arbitrary CAV $i$, a unique identity is assigned to record its crossing order and direction when it enters the CZ. The identity is defined as $(i,d_i,\mathcal{I}_{i}(k))$, where $i\!\in\!\mathcal{N}$ is the crossing order of vehicle $i$ to enter the MZ, $d_i$ is an index denoted as the traveling direction (north, south, east and west), and
$\mathcal{I}_{i}(k)$ is the information set generated by the coordinator at step $k$ by collecting the past optimal state sequences of vehicle $i$:
\begin{equation}
\begin{aligned}
\mathcal{I}_{i}(k) &= [\mathbf{{x}_{i}}(0),\mathbf{{x}_{i}}(1),\cdots,\mathbf{{x}_{i}}(k)] \in \mathbb{R}^{(N_p+1)\times (4 k)},\,
\end{aligned} 
\end{equation}
where $\mathbf{x}_i(k)\!=\![x_i(k|k),x_i(k\!+\!1|k),\ldots,x_i(k\!+\!N_p)]^{\top}$ is the optimal state sequence computed by the local MPC on CAV $i$ at step $k$ within the prediction horizon $N_p$.

Then, based on the identity of each CAV and the information exchange rule shown below, these information sets $\mathcal{I}_{i}(k)$ will be sent to specific CAVs which require them to support their DMPC algorithm. Taking an arbitrary CAV $i$ as an example, the information exchange rule is shown as follows:
\begin{enumerate}
    \item For $\mathcal{C}_{i}\notin \emptyset$, CAV $i$ requires $\mathcal{I}_{h}(k)$ of CAV $h\!\in\! \mathcal{C}_{i}$ ($h\!<\!i$), which is immediately ahead of CAV $i$ in the same lane. 
    \item For $\mathcal{L}_{i}\!\notin\! \emptyset$, CAV $i$ requires $\mathcal{I}_{q}(k)$ of CAV $q\!\in\! \mathcal{L}_{i}$ ($q\!<\!i$), which is the last element (CAV index) in set $\mathcal{L}_{i}$.
    \item For $\mathcal{O}_{i}\!\notin \!\emptyset$, CAV $i$ requires $\mathcal{I}_{p}(k)$ of CAV $p\!\in\!\mathcal{O}_{i}$ ($p\!<\!i$), which is the last element (CAV index) in set $\mathcal{O}_{i}$.
\end{enumerate}

For the decentralized MPC framework, each CAV can achieve the collision avoidance constraints \eqref{eq:TTCapprox}, \eqref{eq:lateral}, \eqref{eq:opposite} based on the information set $\mathcal{I}_{i}(k)$ assigned from the coordinator subject to the information exchange rule. To achieve the rear-end collision avoidance constraints, an arbitrary CAV $i$ requires the information set of the CAV $h$ which is directly ahead of CAV $i$, such that ${h}\!\in\! \mathcal{C}_{i}$. This constraint is reformulated as follows:  %
\begin{equation}
t_{i}(k+j+1|k)-t_{h}(k+j+1|k) \geq \beta(k+j+1|k),
\label{eq:TTCapprox1}    
\end{equation}
where $j\!\in\!\mathbb{N}_{[0,N_p-1]}$, and $t_{h}(k\!+\!j\!+\!1|k)$ is the past information of CAV $h$ saved in the information set $\mathcal{I}_{h}(\bar{k})$, where $\bar{k}\!>\!k$ is the corresponding distance step of CAV $h$ when the $i$th CAV is at step $k$, that is $t_{h}(\bar{k} \Delta s)\!=\!t_{i}(k \Delta s)$.

The reformulation of the lateral collision avoidance constraint \eqref{eq:lateral} between CAV $i$ and CAV $q\in\mathcal{L}_i$ in the decentralized framework is given as below:
\begin{equation}\label{eq:decentralized_lateral}
t_{i}(k+j+1|k) \geq \hat{t}_{{q}}^{\mathcal{L}}(L+S),\,\,\,j\in\mathbb{N}_{[0,N_p-1]}\,,
\end{equation}
where $k+j+1=\alpha_1$ with $\alpha_1\!\in\!\mathbb{N}_{[k,\,k+N_p]}$. The exit time of the ${q}$th CAV of MZ, $\hat{t}_{q}^{\mathcal{L}}(L+S)$, is estimated by:
\begin{equation}\label{eq:decentralize_lateral_right}
\begin{aligned}
&\hat{t}_{{q}}^{\mathcal{L}}\!(L\!+\!S)\!\!=\!\! 
\left\{\begin{array}{lll} 
\hspace{-1mm}t_{{q}}(\bar{k}\!+\!l\!+\!1|\bar{k}),\,\displaystyle\text{if}\,\, l\!<\!N_p,\,\bar{k}\!+\!l\!+\!1\!=\!\alpha, \\
\hspace{-1mm}\vspace{1mm}\displaystyle
t_{{q}}(\bar{k}\!+\!N_p|\bar{k})\!+\!\frac{2(L\!+\!S\!-\!(\bar{k}\!+\!N_p)\Delta s)}{v_{{q}}(\bar{k}\!+\!N_p|\bar{k})\!+\!\bar{v}_i},\\
\hspace{35mm}\displaystyle
\text{if}\,\,\bar{k}\!+\!N_p\!<\!\alpha,
\end{array}\right.
\end{aligned}
\end{equation}
where $l\!\in\!\mathbb{N}_{[0,N_p-1]}$, $\bar{v}_i$ is the desired terminal velocity and $\bar{k}$ has $t_{q}(\bar{k} \Delta s)\!=\!t_{i}(k \Delta s)$, which represents the associated step of the ${q}$th CAV. For the first case in \eqref{eq:decentralize_lateral_right}, since the prediction horizon of the ${q}$th CAV has already covered the terminal distance $L\!+\!S$, $\hat{t}_{q}^{\mathcal{L}}(L\!+\!S)$ can be directly obtained. For the second case in \eqref{eq:decentralize_lateral_right}, the ${i}$th CAV can not directly derive the terminal time $\hat{t}_{q}^{\mathcal{L}}(L+S)$. Therefore, the exit time of the ${q}$th CAV is approximated by the sum of the present predictive time of the last predictive horizon step of CAV ${q}$, $t_{q}(\bar{k}\!+\!N_p|\bar{k})$, and the time such that CAV ${q}$ should spend to finish the rest of the distance with an average speed of its terminal speed of the present horizon and the desired terminal velocity. 

In terms of the no collision condition in the MZ between CAV $i$ and CAV $p\!\in\!\mathcal{O}_{i}$, the constraint of \eqref{eq:opposite} is rewritten in a decentralized manner as below:
\begin{equation}\label{eq:decentralized_FIFO}
\begin{aligned}
&t_{i}(k+j+1|k) \geq \hat{t}_{p}^{\mathcal{O}}(L),\,\,k+j+1=\alpha_1\\
&t_{i}(k+l+1|k) \geq \hat{t}_{p}^{\mathcal{O}}(L+S),\,\,k+l+1\!=\alpha
\end{aligned}
\end{equation}
where $j,l\!\in\!\mathbb{N}_{[0,N_p-1]}$, while $\hat{t}_{p}^{\mathcal{O}}(L)$ and $\hat{t}_{p}^{\mathcal{O}}(L+S)$ can be obtained by applying the same method shown in \eqref{eq:decentralize_lateral_right}.

Therefore, the convex DMPC framework for a single vehicle $i\!\in\! \mathcal{N}$ in space domain is presented as follows:
\begin{subequations}\label{eq:decentralized}
\begin{align}
    & \mathop {\min}\limits_{\mathbf{u}_i} \hspace{3mm}  J_i(\mathbf{x}_i
    (k),\mathbf{u}_i(k)) \label{eq:J_decentralized} \\
    &\textbf{s.t.}:  \,(\,\forall\, k \in [0,\,\alpha],\,j\in\mathbb{N}_{[0,N_p-1]}) \nonumber\\
    &{x}_i(k|k)={x}_i(k) \label{eq:decentralized_initial}\\
    & {x}_i(k\!+\!j\!+\!1|k)\!=\!A_i{x}_i(k\!+\!j|k)\!+\!B_i{u}_i(k\!+\!j|k)\!+\!B_{c,i},  \label{eq:dynamics_decentralized}\\
    & {x}_i(k\!+\!j|k)\!\in\!\mathbb{X}_i,\,{u}_i(k\!+\!j|k)\!\in\!\mathbb{U}_i\,,   \label{eq:constraints_decentralized}  \\
    & \left({x}_i(k\!+\!j|k),{u}_i(k\!+\!j|k)\right)\!\in\!\mathbb{X}_i\times \mathbb{U}_i\,, \nonumber \\
    & \textbf{given.}:  {x}_i(0) \label{eq:initial_decentralized}
\end{align}
\end{subequations}
where $\mathbf{u}_i(k)\!=\![u_i(k|k),u_i(k+1|k),\cdots,u_i(k+N_p-1|k)]^{\top}$,
$x_i(k)$ in \eqref{eq:decentralized_initial} is the actual state measured at step $k$,  
the system dynamics \eqref{eq:dynamics_decentralized} collects \eqref{eq:dynamics_discretize}, the convex set $\mathbb{X}_i$ in \eqref{eq:constraints_decentralized} collects \eqref{eq:Bound_E_i}, \eqref{eq:terminalv}, \eqref{eq:SOC_limits}, \eqref{eq:TTCapprox1}, \eqref{eq:decentralized_lateral}, and \eqref{eq:decentralized_FIFO}, the convex set $\mathbb{U}_i$ in \eqref{eq:constraints_decentralized} collects \eqref{eq:Fb_limits}, \eqref{eq:HEV_FPS_bound}, \eqref{eq:HEV_FSS_bound}, and the convex set $\mathbb{X}_i\times \mathbb{U}_i$ in \eqref{eq:constraints_decentralized} collects \eqref{eq:dt_path}. 
The initial condition \eqref{eq:initial_decentralized} provides initial state (entry state at CZ) of each CAV $x_i(0)$, which satisfies the Assumptions~\ref{ass:entrytime} and \ref{ass:initial}. The objective function \eqref{eq:J_decentralized} in this is designed as a multi-objective function, which is described as below:
\begin{multline}\label{eq:J_HEV}
    J_i= \sum_{j=0}^{N_p-1}\left(W_{1}\zeta_i(k+j|k)\!-\! W_{2}F_{b,i}(k+j|k)\right)\Delta s +\\ W_{3}m_{f,i}(k+N_p|k) \!+\! W_{4}(\mathrm{SOC}(k+N_p|k)\!-\!\mathrm{SOC}(k|k))^2,
\end{multline}
where $W_1$, $W_2$, $W_3$, $W_4$ are the weighting factors. The first three terms aim to minimize the traveling time, and the usage of mechanical braking and the fuel consumption of \eqref{eq:HEV_m_f}, respectively, and the last term is imposed to achieve the battery charge sustaining (CS) condition over the entire prediction horizon.

\section{Numerical Result}
\label{sec:simulation}

This section provides numerical examples to investigate the performance of the proposed convex DMPC strategy for a group of $N(t)\!=\!20$ CAVs at an autonomous intersection crossing. 
Two convex benchmark vehicle models (i.e. CV~\citep{HAN2019558} and BEV~\citep{pan2022optimal}) are considered and formulated within
the same DMPC framework \eqref{eq:decentralized} for fair comparisons.
The investigation contains twofold, 1) the effectiveness and validity of the proposed convex sHEVs model and the DMPC strategy, and 
2) the impact of the fuel and electricity energy economy between sHEVs and the two other vehicle types (i.e., CVs and BEVs) in the context of the electricity and petrol price in the UK over the past 12 months.

The parameters of the intersection in this work are $L \!=\! 150$~m and $S \!=\! 10$~m with a sampling interval $\Delta s\!=\!2$~m.
The prediction horizon length of the DMPC method is defined as $N_p \!=\! 30$, . 
The terminal velocity leaving the MZ is set to be identical for all CAVs at $\bar{v}_i(L+S) \!=\! 10$~m/s for fair and easy comparison in different simulation scenarios. Without loss of generality and credibility of the simulation results, this numerical analysis has applied 10 sets of different initial conditions datasets with an arrival rate of 700 veh/h per lane. Moreover, in each of the datasets, the initial velocity $v_i(0)$ and entrance time $t_i(0)$ for all CAVs are randomly initialized
subject to the constraints imposed in Assumptions~\ref{ass:entrytime} and \ref{ass:initial}. Note that the initial speeds of all CAVs follow the uniform distribution within $[v_{\min}, v_{\max}]$ and the entrance times for all CAVs follow Poisson distribution, respectively. Moreover, the direction and type of each CAV are also randomly generated. 
For simplicity and illustration purposes, the crossing sequence $\mathcal{N}(t)$ in this work is predefined such that all CAVs obey the first-in-first-out policy (i.e., all CAVs enter and leave the MZ in the same order they arrive at the CZ).
The proposed convex DMPC strategy is solved by YALMIP and MOSEK~\citep{Lofberg2004} in MATLAB 2021b on a personal computer with an M1 chip and 8 GB of RAM.
\begin{figure}[t!]
    \centering
    \includegraphics[width=.9\columnwidth]{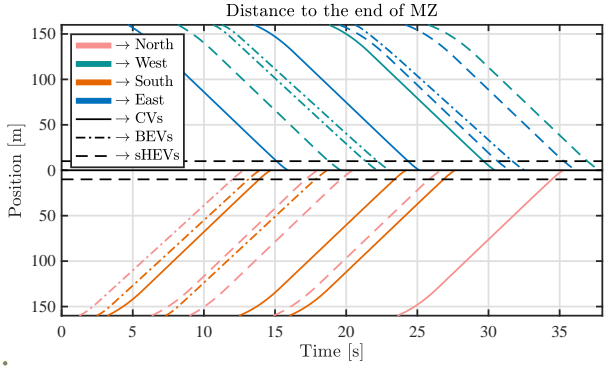}\\ [-2ex]
    \caption{The optimal traveled distance trajectories of 20 CAVs obtained by solving the convex DMPC~\eqref{eq:decentralized} at an arrival rate of 700 veh/h per lane and an average traveled time of approximately 12s.}
    \label{fig:trajectory}
\end{figure}
\begin{figure}[t!]
    \centering
    \includegraphics[width=.9\columnwidth]{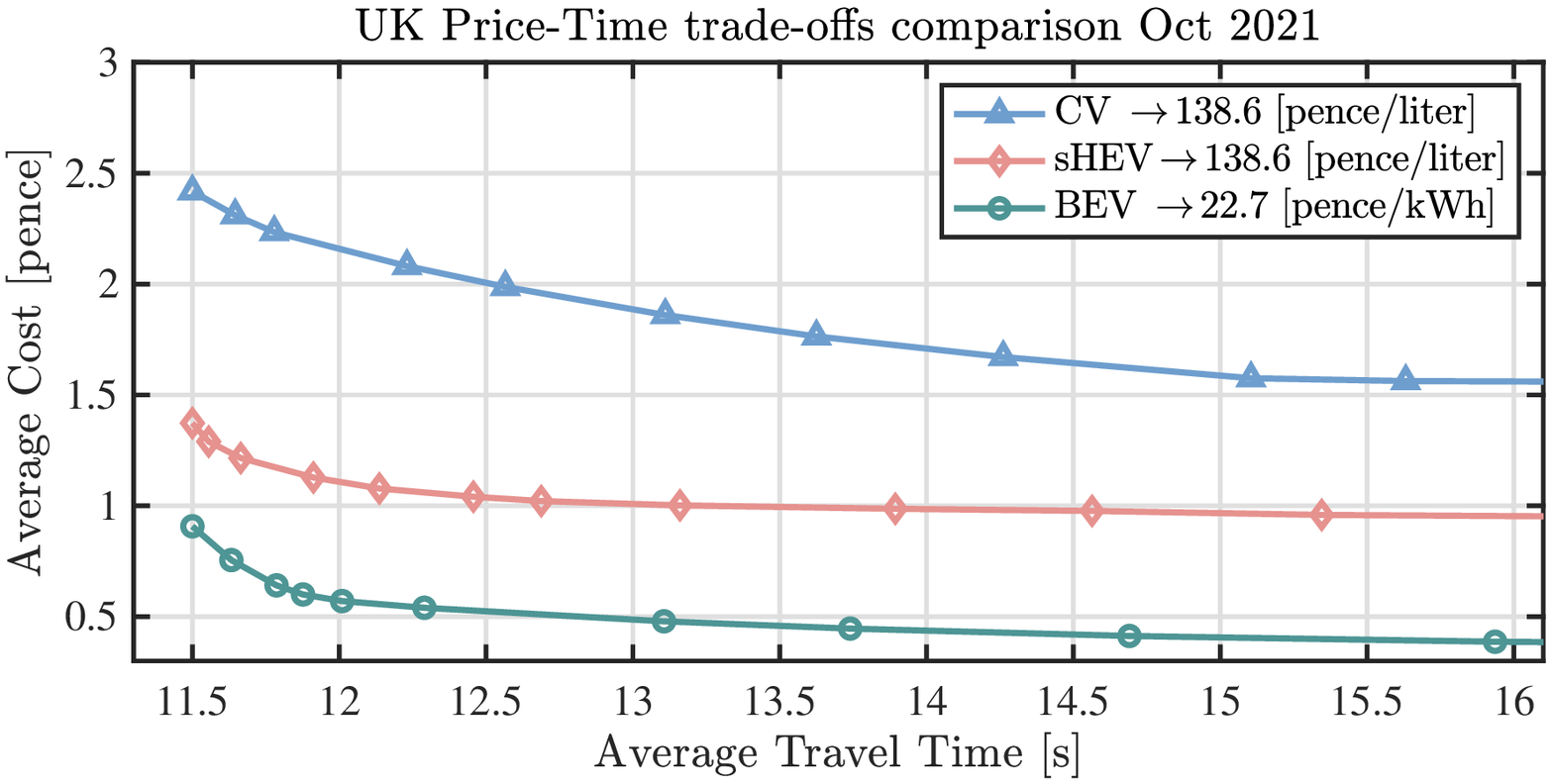} \\[-4ex]
    \includegraphics[width=.9\columnwidth]{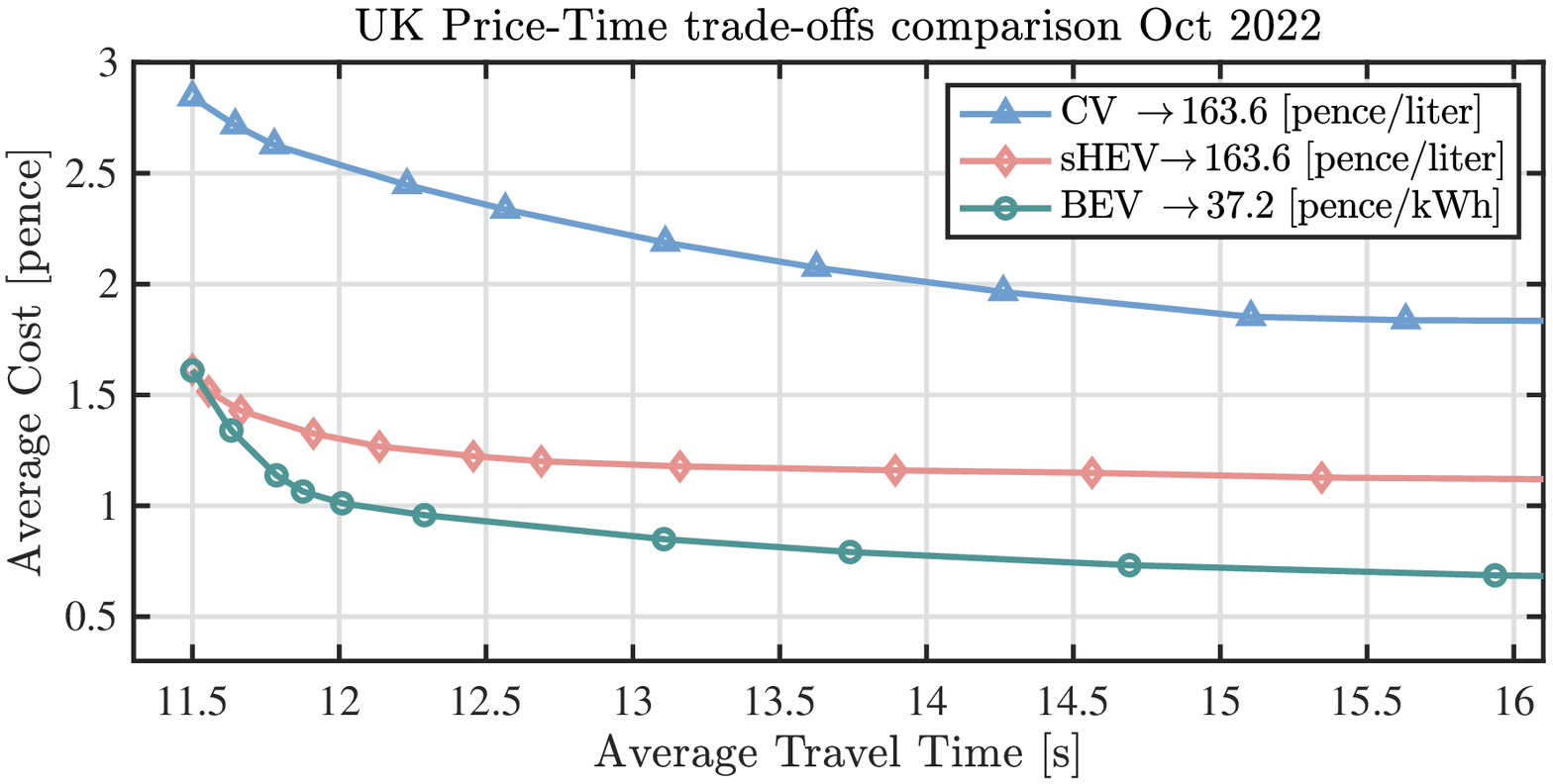} \\ [-2ex]
    \caption{The trade-offs between average price and average travel time for optimization problems with three different vehicle types (i.e CVs, sHEVs, and BEVs) in each case at an arrival rate of 700 veh/h per lane. The average prices are calculated by the price datasets of Oct. 2021 (top) and Oct. 2022 (bottom)~\citep{ukelectricity, ukgasoline}.}
    \label{fig:2122tradeoff}
\end{figure}
\begin{figure}[t!]
    \centering
    \includegraphics[width=1\columnwidth]{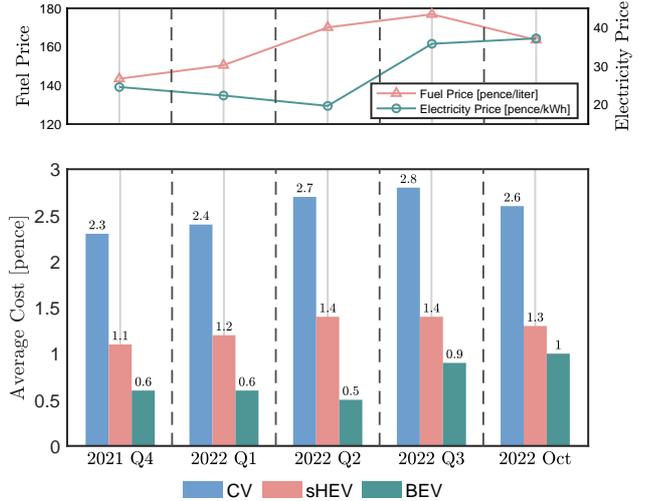} \\ [-2ex]
    \caption{Top: the average price of fuel and electricity in the UK from 2021 Q4, 2022 Q1, 2022 Q2, 2022 Q3, and October 2022. Bottom: the average travel cost paid for three vehicle types at an average time consumption of approximately $12$ s. The average costs are calculated by the price datasets in~\cite{ukelectricity, ukgasoline}.}
    \label{fig:compare}
\end{figure}

In the first instance, the optimal traveled distance trajectories of the 20 CAVs (i.e., 7 CVs, 6 sHEVs, and 7 BEVs) obtained by solving convex DMPC~\eqref{eq:decentralized} at an average traveled time of approximately 12s are presented in Fig.~\ref{fig:trajectory}.
It can be observed that all vehicles follow the cooperatively assigned crossing order, as they leave the CZ in the same order as they enter the CZ. Moreover, none of the trajectories shown with the same color intersect. For two consecutive vehicles traveling in perpendicular directions, the latter one enters the MZ only after the first one leaves the MZ. This result shows that the rear-end and lateral collision avoidance are satisfied for all CAVs, which verifies the effectiveness and validity of the convex sHEVs model and DMPC strategy.

To compare the performance between sHEVs and the other two benchmark vehicle models (i.e., CVs and BEVs) with various electricity and petrol fuel prices,
three autonomous intersection problems with only one vehicle type in each case are formulated under the same proposed DMPC framework~\eqref{eq:decentralized}. 
Fig.~\ref{fig:2122tradeoff} presents the trade-offs of the three autonomous intersection optimization problems for a series of combinations of weighting factors in~\eqref{eq:J_decentralized} between the average energy price and traveled time of all 20 CAVs under ten simulation trials with randomly generated initial conditions. 
Note that petrol fuel and electric energy usages are converted to energy prices in this study based on the datasets of Oct. 2021 and Oct. 2022 obtained from~\citep{ukelectricity, ukgasoline}.
As shown in Fig.~\ref{fig:2122tradeoff}, the average price of CVs is always the highest among the rest cases, which indicates the importance of traffic electrification as it can significantly reduce the average operating cost. 
In terms of sHEVs and BEVs, the gap between each average cost is reduced as the average travel time decreases. Taking the optimal results with the average travel time at 11.5~s as an example, the traveling cost of sHEVs already reaches the same level as BEVs in Oct. 2022, while a 53\% average price gap can be found in the results in Oct. 2021. These results show the advantage of sHEVs as the electricity price increases, especially when the optimization emphasizes more on the minimization of the travel time to achieve efficient transportation.

To further investigate the findings in terms of the reduced average cost between sHEVs and BEVs in Fig.~\ref{fig:2122tradeoff}, a bar chart with detailed average cost from 2021 Q4 to Oct. 2022 at an average time cost of approximately $12$~s is presented in Fig.~\ref{fig:compare}.
The top figure in Fig.~\ref{fig:compare} shows the price of petrol fuel increased by 14\% while the electricity price witnessed a sharp rise of 52\% compared to the 2021 Q4 electricity market prices. 
The corresponding consequences for the vehicle traveling cost are shown at the bottom in Fig.~\ref{fig:compare},
where the gap of the average price between sHEVs and BEVs on operation dropped from 83\% to 30\% from 2021 Q4 to Oct. 2022. Although both CVs and sHEVs consume petrol fuel for operation primarily, the hybrid powertrain can provide higher energy efficiency. Overall, these findings indicate that the sharply increasing electricity price over the last year is weakening the benefit of BEVs, meanwhile highlighting the importance and benefits of sHEVs.

\section{Conclusions}
\label{sec:conclusions}
This paper addressed the signal-free intersection crossing problem for connected and automated vehicles (CAVs) with a series hybrid electric powertrain. The dynamic model of sHEVs is convexified in the space domain. The optimization problems are formulated within a convex decentralized model predictive control (DMPC) framework to ensure a rapid search and unique optimal solution. Numerical examples first validate the effectiveness of the proposed methods with convex relaxation and approximation. 
The performance of the proposed approach is evaluated against the optimal results yielded by solving the same decentralized intersection problems but with two convex benchmark models of conventional vehicles (CVs) and battery electric vehicles (BEVs).
By utilizing the petrol price and electricity charging prices from 2021 Q4 to Oct 2022, it can be found that although the last year witnessed a sharply increasing electric price, CVs are still the most expensive to travel with. Moreover, the average energy price to travel with sHEVs approaches or even sometimes reaches the same level as with BEVs, especially under an efficient transportation strategy (less average traveling time consumption), which highlights the importance and the benefits of sHEVs in the current increasing electricity price market situation in the UK. 

Future work will focus on a more comprehensive analysis of the performance of various HEV types under the turbulent electricity market, including the parallel HEVs, and plug-in electric vehicles for autonomous intersection crossing problem. Robust control strategies will also be developed to reduce and tackle the potential uncertainties in mathematics models, communications, and measurements.

\bibliography{ifacconf}

\begin{thebibliography}{25}
\providecommand{\natexlab}[1]{#1}
\providecommand{\url}[1]{\texttt{#1}}
\providecommand{\urlprefix}{URL }
\expandafter\ifx\csname urlstyle\endcsname\relax
  \providecommand{\doi}[1]{doi:\discretionary{}{}{}#1}\else
  \providecommand{\doi}{doi:\discretionary{}{}{}\begingroup
  \urlstyle{rm}\Url}\fi

\bibitem[{{\v{C}}akija et~al.(2019){\v{C}}akija, Assirati, Ivanjko, and
  Cunha}]{vcakija2019autonomous}
{\v{C}}akija, D., Assirati, L., Ivanjko, E., and Cunha, A.L. (2019).
\newblock Autonomous intersection management: a short review.
\newblock In \emph{2019 International Symposium ELMAR}, 21--26. IEEE.

\bibitem[{Chalaki and Malikopoulos(2022{\natexlab{a}})}]{chalaki2022optimal}
Chalaki, B. and Malikopoulos, A.A. (2022{\natexlab{a}}).
\newblock Optimal control of connected and automated vehicles at multiple
  adjacent intersections.
\newblock \emph{IEEE Transactions on Control Systems Technology}, 30(3),
  972--984.

\bibitem[{Chalaki and Malikopoulos(2022{\natexlab{b}})}]{chalaki2021priority}
Chalaki, B. and Malikopoulos, A.A. (2022{\natexlab{b}}).
\newblock A priority-aware replanning and resequencing framework for
  coordination of connected and automated vehicles.
\newblock \emph{IEEE Control Systems Letters}, 6, 1772--1777.

\bibitem[{Chen et~al.(2019)Chen, Evangelou, and Lot}]{chen2019series}
Chen, B., Evangelou, S.A., and Lot, R. (2019).
\newblock Series hybrid electric vehicle simultaneous energy management and
  driving speed optimization.
\newblock \emph{IEEE/ASME Transactions on Mechatronics}, 24(6), 2756--2767.

\bibitem[{{Department for Business, Energy \& Industrial
  Strategy}(2022)}]{ukgasoline}
{Department for Business, Energy \& Industrial Strategy} (2022).
\newblock {Weekly Prices time series (from 2003)}.
\newblock
  \url{https://www.gov.uk/government/statistics/weekly-road-fuel-prices}.
\newblock Access date: 1 Nov 2022.

\bibitem[{Gholamhosseinian and Seitz(2022)}]{Gholamhosseinian2022}
Gholamhosseinian, A. and Seitz, J. (2022).
\newblock A comprehensive survey on cooperative intersection management for
  heterogeneous connected vehicles.
\newblock \emph{IEEE Access}, 10, 7937--7972.

\bibitem[{Hadjigeorgiou and Timotheou(2022)}]{hadjigeorgiou2022real}
Hadjigeorgiou, A. and Timotheou, S. (2022).
\newblock Real-time optimization of fuel-consumption and travel-time of {CAVs}
  for cooperative intersection crossing.
\newblock \emph{IEEE Transactions on Intelligent Vehicles}.

\bibitem[{Han et~al.(2019)Han, Vahidi, and Sciarretta}]{HAN2019558}
Han, J., Vahidi, A., and Sciarretta, A. (2019).
\newblock Fundamentals of energy efficient driving for combustion engine and
  electric vehicles: An optimal control perspective.
\newblock \emph{Automatica}, 103, 558--572.

\bibitem[{Hult et~al.(2018)Hult, Zanon, Gros, and Falcone}]{Hult:ecc19}
Hult, R., Zanon, M., Gros, S., and Falcone, P. (2018).
\newblock Energy-optimal coordination of autonomous vehicles at intersections.
\newblock In \emph{2018 European Control Conference (ECC)}, 602--607.

\bibitem[{Jian et~al.(2021)Jian, LiJun, and Liang}]{shevJian2021}
Jian, C., LiJun, Q., and Liang, X. (2021).
\newblock Cooperative control of connected hybrid electric vehicles and traffic
  signals at isolated intersections.
\newblock \emph{IET Intelligent Transport Systems}, 14, 1903--1912.

\bibitem[{Liu et~al.(2020)Liu, Zhang, Zhang, Wu, Gao, and Zhang}]{liu2020high}
Liu, C., Zhang, Y., Zhang, T., Wu, X., Gao, L., and Zhang, Q. (2020).
\newblock High throughput vehicle coordination strategies at road
  intersections.
\newblock \emph{IEEE Transactions on Vehicular Technology}, 69(12),
  14341--14354.

\bibitem[{L{\"{o}}fberg(2004)}]{Lofberg2004}
L{\"{o}}fberg, J. (2004).
\newblock Yalmip : A toolbox for modeling and optimization in {MATLAB}.
\newblock In \emph{Proc. of the CACSD Conference}. Taipei, Taiwan.

\bibitem[{Meng et~al.(2018)Meng, Li, Wang, Li, and Li}]{Meng:18}
Meng, Y., Li, L., Wang, F.Y., Li, K., and Li, Z. (2018).
\newblock Analysis of cooperative driving strategies for nonsignalized
  intersections.
\newblock \emph{IEEE Transactions on Vehicular Technology}, 67(4), 2900–2911.

\bibitem[{Mih{\'a}ly et~al.(2020)Mih{\'a}ly, Farkas, and
  G{\'a}sp{\'a}r}]{mihaly2020model}
Mih{\'a}ly, A., Farkas, Z., and G{\'a}sp{\'a}r, P. (2020).
\newblock Model predictive control for the coordination of autonomous vehicles
  at intersections.
\newblock \emph{IFAC-PapersOnLine}, 53(2), 15174--15179.

\bibitem[{Namazi et~al.(2019)Namazi, Li, and Lu}]{namazi2019intelligent}
Namazi, E., Li, J., and Lu, C. (2019).
\newblock Intelligent intersection management systems considering autonomous
  vehicles: A systematic literature review.
\newblock \emph{IEEE Access}, 7, 91946--91965.

\bibitem[{Pan et~al.(2022{\natexlab{a}})Pan, Chen, Dai, Timotheou, and
  Evangelou}]{pan2022TCST}
Pan, X., Chen, B., Dai, L., Timotheou, S., and Evangelou, S.A.
  (2022{\natexlab{a}}).
\newblock A hierarchical robust control strategy for decentralized signal-free
  intersection management.
\newblock \emph{arXiv preprint arXiv:2206.14986}.

\bibitem[{Pan et~al.(2022{\natexlab{b}})Pan, Chen, Timotheou, and
  Evangelou}]{pan2022optimal}
Pan, X., Chen, B., Timotheou, S., and Evangelou, S.A. (2022{\natexlab{b}}).
\newblock A convex optimal control framework for autonomous vehicle
  intersection crossing.
\newblock \emph{IEEE Transactions on Intelligent Transportation Systems}.

\bibitem[{Riegger et~al.(2016)Riegger, Carlander, Lidander, Murgovski, and
  Sjöberg}]{add_centralized}
Riegger, L., Carlander, M., Lidander, N., Murgovski, N., and Sjöberg, J.
  (2016).
\newblock Centralized {MPC} for autonomous intersection crossing.
\newblock In \emph{2016 IEEE 19th International Conference on Intelligent
  Transportation Systems (ITSC)}, 1372--1377.

\bibitem[{Tang et~al.(2021)Tang, Duan, Hu, Pu, Cao, and Lin}]{shevTANG2021}
Tang, X., Duan, Z., Hu, X., Pu, H., Cao, D., and Lin, X. (2021).
\newblock Improving ride comfort and fuel economy of connected hybrid electric
  vehicles based on traffic signals and real road information.
\newblock \emph{IEEE Transactions on Vehicular Technology}, 70(4), 3101--3112.
\newblock \doi{10.1109/TVT.2021.3063020}.

\bibitem[{{TRADING ECONOMICS}(2022)}]{ukelectricity}
{TRADING ECONOMICS} (2022).
\newblock {UK Electricity Spot Prices (GBP/MWh)}.
\newblock
  \url{https://tradingeconomics.com/united-kingdom/electricity-price#:~:text=Electricity%20Price%20in%20the%20United,the%20United%20Kingdom%20Electricity%20Price.}
\newblock Access date: 1 Nov 2022.

\bibitem[{Xiang et~al.(2018)Xiang, Li, Jian-Qiao, and Sun}]{Xiang2018Signal}
Xiang, Li, Jian-Qiao, and Sun (2018).
\newblock Signal multiobjective optimization for urban traffic network.
\newblock \emph{IEEE Transactions on Intelligent Transportation Systems},
  19(11), 3529--3537.

\bibitem[{Yang et~al.(2017)Yang, Rakha, and Ala}]{yang2017cooperative}
Yang, H., Rakha, H., and Ala, M.V. (2017).
\newblock Eco-cooperative adaptive cruise control at signalized intersections
  considering queue effects.
\newblock \emph{IEEE Transactions on Intelligent Transportation Systems},
  18(6), 1575--1585.

\bibitem[{Zhao and Malikopoulos(2022)}]{zhao2022enhanced}
Zhao, L. and Malikopoulos, A.A. (2022).
\newblock Enhanced mobility with connectivity and automation: A review of
  shared autonomous vehicle systems.
\newblock \emph{IEEE Intelligent Transportation Systems Magazine}, 14(1),
  87--102.

\bibitem[{Zhong et~al.(2021)Zhong, Nejad, and Lee}]{zhong2020autonomous}
Zhong, Z., Nejad, M., and Lee, E.E. (2021).
\newblock Autonomous and semiautonomous intersection management: A survey.
\newblock \emph{IEEE Intelligent Transportation Systems Magazine}, 13(2),
  53--70.

\bibitem[{Zhou et~al.(2016)Zhou, Zhang, Li, and Fathy}]{zhou2016pseudospectral}
Zhou, W., Zhang, C., Li, J., and Fathy, H.K. (2016).
\newblock A pseudospectral strategy for optimal power management in series
  hybrid electric powertrains.
\newblock \emph{IEEE Transactions on Vehicular Technology}, 65(6), 4813--4825.

\end{thebibliography}
\end{document}